\documentclass[12pt,fleqn]{article}
\usepackage{verbatim,amssymb,amsmath,graphics,amsthm,amsfonts}
\numberwithin{equation}{section}

\def\<{\langle}
\def\>{\rangle}
\def\+{\phi^+}
\def\-{\psi^-}

\def\b{\begin{equation}}

\def\e{\end{equation}}

\def\E{{\cal E}}
\def\F{{\cal F}}

\def\H{{\cal H}}

\def\S{{\cal S}}

\begin{document}
\title{On the applications of Hardy class functions in scattering
theory}

\author{M.~Gadella\\
Department of Theoretical Physics, Atomic Physics and Optics\\
The University of Valladolid, Valladolid, Spain\\
\\
S.~Wickramasekara\\
Department of Physics, Grinnell College, Grinnell, IA 50112}

\maketitle

\begin{abstract}
This paper is a response to an article~\cite{R} recently published
in Journal of Physics A: Mathematical and Theoretical. The article
claims that the theory of resonances and decaying states based on
certain rigged Hilbert spaces of Hardy functions is physically
untenable. In this paper we show that all of the  key conclusions of
\cite{R} are the result of either the errors in mathematical
 reasoning
 or an inadequate understanding of the literature on the subject.
\end{abstract}

\section{Introduction}\label{sec1}

A recently published work \cite{R} carries out an  analysis of time
asymmetric quantum theory (TAQT) of resonance scattering and decay.
This theory is anchored in a pair of rigged Hilbert spaces of Hardy
functions. The primary claim of \cite{R} is that Hardy functions are
incompatible with quantum mechanics.

The author of \cite{R} arrives at this conclusion by solving the
Lippmann-Schwinger equations for a spherical shell potential.
Specifically, the author constructs a class of solutions and argues
that they are not of Hardy class by claiming that they diverge at
infinity. However, the author makes a rudimentary
 mathematical
error here. What the author actually establishes is an upper bound
for the wavefunctions, $|\varphi(E)|\leq f(E)$, where the upper
bound $f(E)$ has an exponential blow up when $E$ approaches infinity
in the
 complex plane.
The author confuses this exponential blow up of the upper bound with
that of the wavefunction itself and it is this rudimentary mistake
that
 leads
the author to conclude that the solutions obtained in \cite{R} are
not
 of
Hardy class.

Equally importantly, even if the author had handled the problem with
 care and
found a class of solutions that are not of Hardy class, the
existence
 of such solutions
certainly does not allow one to conclude that there exist no Hardy
 solutions
to the particular potential studied, let alone all other potentials.
 Therefore, even if the
study of \cite{R} had been conducted with proper mathematical rigor,
it
 could not
have led a conclusion about the veracity of TAQT.

The author of \cite{R} also quotes fragments of statements from the
 literature which,
when viewed with a diffused focus and out of context, lead to hasty,
erroneous conclusions. For instance,  the author's conclusion that
only the zero function has an asymmetric, semigroup time evolution

is the result of such a misrepresentation of the literature.

In this paper, we will explicitly point out the mathematical errors
in \cite{R} and also show how a pair of rigged
 Hilbert
spaces of Hardy functions may be constructed for the spherical shell
 potential
using the techniques of \cite{R} itself.  We will also explicate the
context and the true content of some of the statements by the
proponents of TAQT that the author uses to support the claims of
\cite{R}.

\section{Rigged Hilbert Spaces of Hardy Functions}\label{sec2}

In this section, we present a brief review of the essential
mathematical features of time asymmetric quantum theory. Our intent
here is to give just enough detail so that claims of \cite{R} can be
juxtaposed with the mathematical framework of TAQT. More complete
and systematic presentations of both mathematical and physical
features of TAQT can be found, for instance, in \cite{BG,BGM,CG}.
These expand on the ideas of the earliest publications on the
subject \cite{BIII, BII, GI}.

Time asymmetric quantum theory is based on a pair of rigged Hilbert
spaces constructed from Hardy class functions:
\begin{subequations}
\label{2.1}
\begin{equation}
\tag{\ref{2.1}+} \left.\S\cap\H_+^2\right|_{{\mathbb{R}}^+}\subset
L^2({\mathbb{R}}^+, dE)\subset
\left(\left.\S\cap\H_+^2\right|_{{\mathbb{R}}^+}\right)^\times\label{2.1+}
\end{equation}
\begin{equation}
\tag{\ref{2.1}$-$} \left.\S\cap\H_-^2\right|_{{\mathbb{R}}^+}\subset
L^2({\mathbb{R}}^+, dE)\subset
\left(\left.\S\cap\H_-^2\right|_{{\mathbb{R}}^+}\right)^\times\label{2.1-}
\end{equation}
\end{subequations}
In \eqref{2.1}, the spaces $\H_+^2$ and $\H_-^2$ are Hardy class
functions on the upper and lower half complex planes, respectively.
The Hilbert space $L^2({\mathbb{R}}^+, dE)$ is the norm completion
of wavefunctions in the energy representation. $\S$ is the Schwartz
space on $\mathbb{R}$ (smooth functions that vanish at infinity
faster than the inverse of any polynomial). The symbol
$\left.\right|_{{\mathbb{R}}^+}$ indicates restrictions of functions
of $\S\cap\H_\pm^2$, whose domain is the entire real line, to the
positive semi-axis, $\mathbb{R}^+$. The dual spaces
$\left(\left.\S\cap\H_\pm^2\right|_{{\mathbb{R}}^+}\right)^\times$
 consist
of continuous antilinear functionals on
$\left.\S\cap\H_\pm^2\right|_{{\mathbb{R}}^+}$.

The spaces $\S\cap\H_\pm^2$ can be obtained as the Fourier
transforms of $\S({\mathbb{R}}^\mp)$, the spaces of Schwartz
functions with supports in the half-line $\mathbb{R}^\mp$. That is,
$\S\cap\H_\pm^2=\F[\S({\mathbb{R}}^\mp)]$. This identity is an
immediate implication of a Paley--Wiener theorem \cite{K,KI}, which
states that the Fourier transformation $\F$ is a unitary mapping
between the Hilbert spaces $L^2({\mathbb{R}}^\mp)$ and the Hardy
spaces $\H_\pm^2$, $\F[L^2({\mathbb{R}}^\mp)]=\H_\pm^2$, and the
fact that $\F$ is a homeomorphism on the Schwartz space $\S$. Since
$\S({\mathbb{R}}^\pm)$ are closed subspaces with respect to the
usual nuclear Fr\'echet topology of $\S$, it follows that
$\S\cap\H_\pm^2$ are also nuclear Fr\'echet spaces.

Using a theorem of van Winter \cite{VW}, it can be shown that there
exist one-to-one and onto mappings $\theta_\pm$ between the
functions of $\S\cap\H_\pm^2$ and their restrictions to
$\mathbb{R}^+$, $\left.\S\cap\H_\pm^2\right|_{\mathbb{R}^+}$ (see
\cite{BG} and Appendix of \cite{BMLG}). These mappings can be used
to
 transport
the topology of the spaces $\S\cap\H_\pm^2$ to the spaces
$\left(\left.\S\cap\H_\pm^2\right|_{\mathbb{R}^+}\right)$. Since
$\S\cap\H_\pm^2$ are nuclear Fr\'echet spaces, the spaces
$\left(\left.\S\cap\H_\pm^2\right|_{\mathbb{R}^+}\right)$ are also
nuclear Fr\'echet spaces under this transported topology.
Furthermore, the denseness of $\S\cap\H_\pm^2$ in $\H_\pm^2$ and a
result of van Winter that asserts the denseness of the restrictions
$\left.\H_\pm^2\right|_{{\mathbb{R}}^+}$ in $L^2({\mathbb{R}}^+)$
can be used to prove that the spaces
$\left(\left.\S\cap\H_\pm^2\right|_{\mathbb{R}^+}\right)$ are dense
in the Hilbert space $L^2({\mathbb{R}}^+,dE)$. Therefore, if we
denote by
$\left(\left.\S\cap\H_\pm^2\right|_{\mathbb{R}^+}\right)^\times$ the
spaces of continuous antilinear functionals on
$\left(\left.\S\cap\H_\pm^2\right|_{\mathbb{R}^+}\right)$, then
\eqref{2.1} are a pair of rigged Hilbert spaces.

The Hilbert space $L^2({\mathbb{R}}^+, dE)$ is thus the starting
point of the construction of the rigged Hilbert spaces \eqref{2.1}.
This Hilbert space is what provides the realization of the system in
the energy representation. To be specific, let $\H$ be an abstract
Hilbert space in which there exists a self-adjoint Hamiltonian $H$.
Let $H$ have a non-degenerate, absolutely continuous spectrum that
coincides with the positive real line, $\mathbb{R}^+$, and let the
point and singularly continuous spectra of $H$ be empty. (In fact,
it is not necessary to assume that the point spectrum of $H$ be
empty \cite{BG}.) Then, the spectral representation theorem
\cite{RS} says that there exists a unitary operator $W$, not
necessarily unique, from $\cal H$ to $L^2({\mathbb R}^+, dE)$ such
that $WHW^{-1}$ is the multiplication operator $\E$ in $L^2({\mathbb
R}^+, dE)$. This means that for any $\varphi(E)\in L^2({\mathbb
R}^+, dE)$ with $E\varphi(E)\in L^2({\mathbb R}^+, dE)$, one has
\begin{equation}
(WHW^{-1}\varphi)(E)=(\E\varphi)(E)=E\varphi(E)\label{2.2}
\end{equation}

The operator $W$ that furnishes the spectral representation of $\H$
is not necessarily unique, and in the construction of the rigged
Hilbert spaces \eqref{2.1}, two such operators $W_\pm$ have been
used~\cite{BG,CG}. The existence the operators $W_\pm$ follows from
the spectral representation theorem and the conditions on the
interaction potential that ensures the existence of M{\o}ller wave
operators. In particular, let the Hamiltonian that determines the
time evolution of the system be of the form $H=H_0+V$, where $H_0$
is an operator that governs some suitable ``free dynamics'' and $V$,
the ``interaction potential'' that represents a perturbation to
$H_0$. As is customary, let us assume that both $H$ and $H_0$ have
empty point and singularly continuous spectra and that their
absolutely continuous spectra are non-degenerate and coincide with
the positive real line, $\mathbb{R}^+$. This assumption leads to
notational simplicity but does not limit the generality of the
results.

Under certain conditions on the interaction potential $V$
\cite{RSIII}, there exist unitary M{\o}ller wave operators
$\Omega_\pm$ on $\H$:
\begin{equation}
\Omega_\pm=\displaystyle{\text{s-lim}_{t\to\mp\infty}}e^{iHt}e^{-iH_0t}\label{2.3}
\end{equation}
where the limit s-$\lim$ is the strong limit with respect to the
norm topology of $\H$, the abstract Hilbert space in which the
operators $H$ and $H_0$ are defined. The M{\o}ller operators
$\Omega_\pm$ fulfill the following intertwining relation:
\begin{equation}
H\Omega_\pm=\Omega_\pm H_0\label{2.4}
\end{equation}
Now, by the spectral representation theorem, there exists a unitary
operator $U$ from $\H$ to $L^2({\mathbb{R}}^+,dE)$ such that the
free Hamiltonian $H_0$ is mapped to the multiplication operator,
$UH_0U^{-1}=\E$, in $L^2({\mathbb{R}}^+,dE)$. That is, for all
$\varphi\in L^2({\mathbb{R}}^+,dE)$ such that $\E\varphi\in
L^2({\mathbb{R}}^+)$,
\begin{equation}
\left(UH_0U^{-1}\varphi\right)(E)=\left(\E\varphi\right)(E)=E\varphi(E)\label{2.5}
\end{equation}

From \eqref{2.4} and \eqref{2.5}, it follows that the operators
$W_\pm:=U\Omega_\pm^{-1}$ provide unitary mappings between $\H$ and
$L^2({\mathbb{R}}^+,dE)$ such that the exact Hamiltonian $H$ is
mapped to the multiplication operator $\E$ in
$L^2({\mathbb{R}}^+,dE)$. Thus, for all $\varphi\in
L^2({\mathbb{R}}^+,dE)$ such that $\E\varphi\in
L^2({\mathbb{R}}^+,dE)$,
\begin{equation}
\left(W_\pm
HW_\pm^{-1}\varphi\right)(E)=\left(\E\varphi\right)(E)=E\varphi(E)\label{2.6}
\end{equation}

By taking the inverse image of \eqref{2.1+} under $W_+$ and of
\eqref{2.1-} under $W_-$, we can obtain two rigged Hilbert spaces
whose spectral realizations are given by \eqref{2.1}. Defining
$\Phi_\pm:=W^{-1}_\pm\left(\left.\S\cap\H_\pm^2\right|_{\mathbb{R}^+}\right)$
and letting $\Phi_\pm^\times$ be the (anti)dual spaces of
$\Phi_\pm$, we then have a pair of abstract rigged Hilbert spaces
\begin{equation}
\Phi_\pm\subset\H\subset\Phi_\pm^\times\label{2.7}
\end{equation}
The construction can be summarized by the following diagram:

\begin{equation}
\begin{array}{lllllllllll}
& {\Phi }_{\pm } &  & \subset  &  & \mathcal{H} & & \subset  &  &
{\Phi }_{\pm } ^{\times } &
\\
&  \left\downarrow{{W_\pm}}
\begin{array}{l}
\\
\\
\end{array}
\right.  &  &  &  & \left\downarrow{{W_\pm}}
\begin{array}{l}
\\
\\
\end{array}
\right.  &  &  &  & \;\;\left\downarrow{{W^\times_\pm}}
\begin{array}{l}
\\
\\
\end{array}
\right.  &  \\
& \left. S\cap \mathcal{H}_{\pm }^2\right| _{\mathbb{R}^{+}} &  &
\subset &  & L^2(\mathbb{R}^{+}, dE) &  & \subset  &  & \left(
\left. S\cap
\mathcal{H}_{\pm}^2\right| _{\mathbb{R}^{+}}\right) ^{\times } &  \\
&  \left\downarrow{\theta}^{-1} _{\pm}
\begin{array}{l}
\\
\\
\end{array}
\right.  &  &  &  &  &  &  &  & \;\;\;\left\downarrow\left(
{\theta}^{-1} _{\pm}\right) ^{\times }
\begin{array}{l}
\\

\end{array}
\right.  &  \\
& S\cap \mathcal{H}_{\pm }^2 &  & \subset  &  & L^2(\mathbb{R}, dE)
& & \subset  & & \left( S\cap \mathcal{H}_{\pm }^2\right) ^{\times }
&
\end{array}\label{2.8}\;
\end{equation}

The rigged Hilbert spaces \eqref{2.1} or, equivalently, \eqref{2.7},
have the following important properties:

\begin{itemize}
\item[{2.1.}]\label{item2.1)}
The multiplication operator $\E$ is reduced by both spaces
$\S\cap{\cal H}_\pm^2\Big|_{{\mathbb R}^+}$, i.e.,
\begin{equation}
 {\cal E}\left(  \S\cap{\cal H}_\pm^2\Big|_{{\mathbb R}^+}\right)
 \subset
 \S\cap{\cal H}_\pm^2\Big|_{{\mathbb R}^+}\label{2.9}
\end{equation}
Further, as an operator in $L^2({\mathbb{R}}^+,dE)$ defined with
domain $\S\cap{\cal H}_+^2\Big|_{{\mathbb R}^+}$ or $\S\cap{\cal
H}_-^2\Big|_{{\mathbb R}^+}$, $\E$ is essentially self-adjoint.
Therefore, the nuclear spectral theorem, the precise mathematical
statement for Dirac's continuous basis vector expansion, holds for
$\E$.

By the Stone-von Neumann theorem, the operator $-i\bar{\E}$
generates the one parameter unitary group $U(t)=e^{-i\bar{\E} t}$ in
$L^2({\mathbb{R}^+}, dE)$.

As an operator in $\S\cap{\cal H}_\pm^2\Big|_{{\mathbb R}^+}$, $\E$
is continuous with respect to the nuclear Fr\'echet topology of
these spaces~\cite{BG,CG}. Furthermore, the operators $\pm i\E$
generate two differentiable one parameter {\em semigroups}
$U_\pm(t)=e^{\pm i\E t}$, $t\geq0$, in $\S\cap{\cal
H}_\pm^2\Big|_{{\mathbb R}^+}$.

Since the spaces $\left.\S\cap\H_\pm^2\right|_{{\mathbb{R}}^+}$ of
\eqref{2.1} and $\Phi_\pm$ of \eqref{2.7} are homeomorphic, the
above conclusions for the operator $\E$ in
$\left.\S\cap\H_\pm^2\right|_{{\mathbb{R}}^+}$ also hold for the
operator $H$ in $\Phi_\pm$. In particular, as an operator in the
Hilbert space $\H$, $H$ is essentially self-adjoint in the domains
$\Phi_\pm$ and is reduced by them. As an operator in $\Phi_\pm$, $H$
is continuous and the operators $\pm iH$ generate two one parameter
semigroups $e^{\pm iHt}$ in $\Phi_\pm$. Specifically, let us use the
mappings $W_\pm:{\Phi}_\pm\longmapsto {\cal S}\cap{\cal
H}_\pm^2\Big|_{{\mathbb R}+}$ as defined above for
 $\phi_\mp\in{\Phi}_\pm$
and write $W_\pm\phi_\mp=\varphi_\mp$. Then, for any
$\phi_\mp\in{\Phi}_\pm$
\begin{eqnarray}
W_\pm \, e^{\pm itH}\,\phi_\mp= W_\pm \, e^{\pm
itH}\,W^{-1}_\pm\,W_\pm\,\phi_\mp =e^{\pm it{\cal
E}}\,\varphi_\mp\label{2.9a}
\end{eqnarray}
Since the inclusion $e^{\pm i{\cal
 E}t}\left(\left.\S\cap\H_\pm^2\right|_{\mathbb{R}^+}\right)\subset
\left.\S\cap\H_\pm^2\right|_{\mathbb{R}^+}$ holds if and only if
 $t\geq0$, then it follows from
\eqref{2.9a}
\begin{equation}
e^{\pm itH}\left({\Phi}_\pm\right)\subset\Phi_\pm\quad\text{if and
only
 if}\ t\geq0\label{2.9b}
\end{equation}
That is, the operators $\pm iH$ generate two one parameter
semigroups
 $e^{\pm iHt},\ t\geq0,$ in
$\Phi_\pm$.

\item[{2.2.}]\label{item2.2}
The nuclear spectral theorem holds for $H$ in $\Phi_\pm$. That is,
there exists vectors $|E^-\>\in\Phi_+^\times$ and
$|E^+\>\in\Phi_-^\times$ , $E\in{\mathbb{R}}^+=\text{spectrum}(H)$,
such that for any $\psi^-\in\Phi_+$ and for any $\phi^+\in\Phi_-$,
\begin{eqnarray}
|\psi^-\>&=&\int_0^\infty dE|E^-\>\<^- E|\psi^-\>\nonumber\\
|\phi^+\>&=&\int_0^\infty dE|E^+\>\<^+ E|\phi^+\>\label{2.10}
\end{eqnarray}

The vectors $|E^\pm\>$ are eigenvectors of $H^\times$
\begin{equation}
H^\times|E^\pm\>=E|E^\pm\>\label{2.11}
\end{equation}
Furthermore, by construction of the spaces by means of the unitary
operators $V_\pm:=U\Omega_\pm^{-1}$, it follows that
\begin{equation}
|E^\pm\>=\Omega_\pm^\times|E\>, \quad\text{where}\quad
H_0^\times|E\>=E|E\>\label{2.12}
\end{equation}
In analogy to the heuristic eigenkets of the Lippmann-Schwinger
equations, we call the basis vectors $|E^\pm\>$ that fulfill
\eqref{2.10} -- \eqref{2.12} Lippmann-Schwinger kets.

\item[{2.3.}]\label{item2.3}
The functions in $\left.\S\cap\H_\pm^2\right|_{\mathbb{R}^+}$, dense
subspaces of $L^2({\mathbb{R}^+},dE)$, are supported on
$\mathbb{R}^+$, the spectrum of the physical Hamiltonian $H$.
However, it is a property of Hardy class functions \cite{VW}   that
their values on the entire real line as well as the relevant open
complex half plane are determined by their values on the positive
semi-axis $\mathbb{R}^+$. This property allows the domains of the
functions of $\left.\S\cap\H_\pm^2\right|_{\mathbb{R}^+}$ to be
extended to the entire open half planes ${\mathbb{C}}^\pm$ as well
as to the negative semi-axis $\mathbb{R}^-$. As a result, Gamow
vectors  can be defined as generalized eigenvectors $|z_R^-\>$ of
the Hamiltonian with complex eigenvalue $z_R$, the complex resonance
pole position of the $S$-matrix.

\item[{2.4.}]\label{item2.4} As a
consequence of paragraph \ref{2.1}, the Gamow vectors $|z_R^-\>$,
the Lippmann-Schwinger vectors $|E^\pm\>$, the prepared state
vectors $\phi^+$ and observable vectors $\psi^-$ all have
asymmetric, semigroup time evolutions. It is in this sense that the
rigged Hilbert space theory of Hardy functions is a time asymmetric
quantum theory of resonance scattering and decay.
\end{itemize}

Notice that the construction of the rigged Hilbert spaces
\eqref{2.1} and \eqref{2.7} holds for any potential $V$ that
satisfies the following two conditions:
\begin{enumerate}
\item The full Hamiltonian $H=H_0+V$ is essentially self-adjoint
and has an absolutely continuous spectrum bounded from below.
\item The M{\o}ller wave operators exist and are asymptotically
complete.
\end{enumerate}
{\em Therefore, the assertion that {\rm`nobody has found a potential
to which such a theory applies'} in the article \cite{R} is absurd.}
In fact, as we will see in Section \ref{sec3}, the author of
\cite{R} has provided an example of a potential which is completely
consistent with the Hardy space theory.

\section{Spherical shell potential}\label{sec3}

All of the conclusions of \cite{R} about time asymmetric quantum
theory are drawn from the single example of scattering off a
spherical shell potential. We will show here that this potential is
in fact perfectly consistent with TAQT and that the conclusions of
\cite{R} are the consequences of rudimentary mathematical errors.

Let us first summarize the solutions of the Lippmann-Schwinger
equations obtained in \cite{R} for the spherical shell potential.
The analysis of \cite{R} begins with the radial Schr\"odinger
equation for $l=0$,
\begin{equation}
\left(-\frac{d^2}{dr^2}+V(r)\right)\chi^\pm(r,E)=E\chi^\pm(r,E)\label{3.1}
\end{equation}
where the potential $V(r)$ is given by
\begin{equation}
    V(r)=\left\{\begin{array}{ccc}
      \infty &  & r<0 \\[2ex]
      0 &  & 0\le r\le a \\[2ex]
      1 &  & a\le r\le b  \\[2ex]
      0 & & b<r\\
    \end{array}\label{3.2}
    \right.
\end{equation}
The Lippmann-Schwinger functions are the solutions of the radial
Lippmann-Schwinger equations
\begin{equation}
    \langle r|E^\pm\rangle=\langle r|E\rangle+ \langle
    r|\frac{1}{E-H_0\pm i0}\,V|E^\pm\rangle\label{3.3}
\end{equation}
For $l=0$, the solutions of equation \eqref{3.3} can be obtained as
solutions of the radial Schr\"odinger equation
\begin{equation}
    \left( -\frac{d^2}{dr^2}+V(r)
    \right)\langle r|E^\pm\rangle=E\langle r|E^\pm\rangle\label{3.4}
\end{equation}
subject to the following boundary conditions
\begin{eqnarray}
 &&  \langle r|E^\pm\rangle= 0\, \;\;{\rm if}\quad r\leq0\,. \nonumber
 \\
 &&  \langle r|E^\pm\rangle \quad {\rm is \; continuous \; at} \;
r=a,b\,. \nonumber \\
  && \frac{d}{dr}\,\langle r|E^\pm\rangle \quad  {\rm is \; continuous
\; at} \; r=a,b\,.\nonumber  \\
 &&  \langle r|E^\pm\rangle \sim e^{\mp ikr}-S(E)\,e^{\pm ikr} \quad
{\rm as} \;\; r\longmapsto\infty, \label{3.5}
\end{eqnarray}
where
\begin{equation}
    k:=\sqrt{\frac{2mE}{\hbar^2}}\ \text{and}\
E\in[0,\infty)\label{3.6}
\end{equation}

The analyticity properties of $\langle r|E^\pm\rangle$ are studied
in \cite{RI}. The notation $\langle r|E^\pm\rangle$ suggests that
both $|r\rangle$ and $|E^\pm\rangle$ are functionals on some
suitably defined vector spaces (e.g., test function spaces) and that
$\langle r|E^\pm\rangle$ is an integral kernel. Thus, with the
choice of test function space
\begin{equation}
    {\Phi}:= {\cal S}({\mathbb R}^+/ \{ a,b\} )\label{3.7}
\end{equation}
i.e., the subspace of Schwartz functions supported on the positive
semiaxis ${\mathbb R}^+$ that vanish at the points $a$ and $b$, the
kets $|E^\pm\rangle$ are defined as antilinear functionals by means
of the following formula:
\begin{equation}
 \varphi^\pm(E):= \langle E^\pm|\varphi^\pm\rangle=
 \int_0^\infty dr\ \varphi^\pm(r)
    {\langle E^\pm|r\rangle}=
 \int_0^\infty dr\ \varphi^\pm(r)
    \overline{\langle r|E^\pm\rangle}
    \label{3.8}
\end{equation}
These functions $\varphi^\pm$ are claimed to have analytic
extensions into the complex plane \cite{R}. These extensions in turn
follow from the analytic extensions of $|E^\pm\>$ so that
\begin{equation}
\varphi^+(z)=\int_0^\infty dr \varphi^+(r)\, \langle
z^+|r\rangle\label{3.9}
\end{equation}
Using the following inequality for $\left|\<z^+|r\>\right|$
attributed to \cite{Taylor},
\begin{equation}
\left|\<z^+|r\>\right|\leq C\, \frac{|z|^{1/4}\,r}{1+|z|^{1/2}\,r}
\, e^{|{\rm Im}\,\sqrt{z}|\,r},\label{3.10}
\end{equation}
where $C$ is a positive constant, and another inequality for Jost
 functions
${\mathcal{J}}_-(z)$ drawn from \cite{RI},
\begin{equation}
\left|\frac{1}{{\cal{J}}_-(z)}\right|\leq C,\quad
z\in{\mathbb{C}},^-\label{3.11}
\end{equation}
the author of \cite{R} concludes that for an infinitely
differentiable $\varphi^+(r)$ with behavior
\begin{subequations}
\label{3.12}
\begin{equation}
\tag{\ref{3.12}a} \varphi^+(r)\sim e^{-\frac{r^2}{a}}\ \text{as}\
r\rightarrow\infty,\label{3.12a}
\end{equation}
the function $\varphi^+(z)$ blows up exponentially on $\mathbb{C}$:
\begin{equation}
\tag{\ref{3.12}b} \varphi^+(z)\sim e^{\frac{|{\rm Im}(z)|^b}{b}}\
\text{as}\ z\rightarrow\infty.\label{3.12b}
\end{equation}
\end{subequations}
In \eqref{3.12},  $a$ and $b$ are two real numbers fulfilling
$\frac{1}{a}+\frac{1}{b}=1$. Furthermore, if $\varphi^+(r)$ is an
infinitely differentiable function with compact support, the author
obtains the inequality \cite{R}
\begin{equation}
|\varphi^+(z)|\le C\, \frac{|z|^{1/4}\,A}{1+|z|^{1/2}\,A} \,
e^{|{\rm Im}\,\sqrt{z}|\,A}\label{3.13}
\end{equation}
where $A$ and $C$ are positive constants. From this the author of
\cite{R} concludes that ``when $\varphi^+(r)\in C_0^\infty$,
$\varphi^+(z)$ blows up {\em exponentially} in the infinite arc of
${\mathbb{C}}_{II}^-$":
\begin{equation}
\text{If} \left|\varphi^+(r)\right|=0\ \text{when}\ r>A, \
\text{then}\ |\varphi^+(z)|\sim e^{A\left| {\rm
Im}{\sqrt{z}}\right|}\ \text{as}\ z\rightarrow\infty\label{3.14}
\end{equation}

This conclusion is simply wrong. Even if the mathematics leading to
\eqref{3.13} and \eqref{3.10} is correct, all that one can conclude
from these inequalities is that {\em a particular upper bound} on
$\varphi^+(z)$ blows up for $z\rightarrow\infty$. The inequalities
\eqref{3.13} and \eqref{3.10} do not say anything at all about the
behavior of $\varphi^+(z)$ for $z\rightarrow\infty$ except that
$\left|\varphi^+(z)\right|$ is {\em smaller} than or equal
 to
a function that increases without bound for $z\rightarrow\infty$.
Though trivial, it needs to be pointed out that it is a {\em
divergent lower bound} for $|\varphi^+(z)|$ that proves the
divergence of $\varphi^+(z)$ at infinity. It should also be pointed
out that even if one succeeds in finding a
 basis for
$L^2({\mathbb{R}}^+,dE)$ that does not belong to Hardy spaces,  this
does not mean that one has ruled out  the Hardy solutions. The
spaces $\left.\S\cap\H_\pm^2\right|_{{\mathbb{R}}^+}$ are dense
subspaces of $L^2({\mathbb{R}}^+,dE)$, and so one always has Hardy
functions if one works with a Hilbert space.

In fact, while a better bound that shows the finiteness of
$\left|\varphi^+(z)\right|$ at infinity may not be found easily, it
is
 easy to
show that a rigged Hilbert space of Hardy functions can be
constructed from the solutions of \eqref{3.4} for the spherical
shell potential. To that end, consider the Hilbert spaces
$L^2({\mathbb R}^+,dr)$ and $L^2({\mathbb R}^+,dE)$. Both are the
same Hilbert space of square integrable functions on the positive
semiaxis, although we use a different notation for each because the
former is the space of radial wavefunctions and the latter contains
the same states in the energy representation. The mappings $U_\pm$
\begin{equation}
 U_\pm: \;L^2({\mathbb R}^+,dr)\longmapsto L^2({\mathbb
R}^+,dE)\label{3.15}
\end{equation}
given by
\begin{equation}
(U_\pm f)(E):=  \widehat f^\pm(E)= \int_0^\infty
  f(r)\,\langle r|E^\pm\rangle \,dr,\quad f\in
L^2({\mathbb{R}}^+,dr)\label{3.16}
\end{equation}
are unitary. The inverse operators $U_\pm^{-1}$ of  \eqref{3.15} are
given by
\begin{equation}
\left(U_\pm^{-1}\widehat f^\pm\right)(r)= f(r)= \int_0^\infty
\widehat f^\pm(E)\,\overline{\langle
    r|E^\pm\rangle}\, dE\label{3.17}
\end{equation}
Here,  $\<r|E^\pm\>$ are the solutions of \eqref{3.4}.

Now, let  $\widehat\varphi^+\in {\cal S}\cap{\cal
H}_+^2\Big|_{{\mathbb R}^+}$. It is obvious that
$\widehat\varphi^+\in L^2({\mathbb R}^+,dE)$ and therefore we can
use it in the integrand of \eqref{3.17} to obtain a function
$\varphi^+=U_+^{-1}\widehat\varphi^+\in L^2({\mathbb R}^+,dr)$.
Since $U_+$ is unitary, the use of $\varphi^+$ in \eqref{3.16}
returns the original function
$\widehat\varphi^+\in\left.\S\cap\H_-^2\right|_{\mathbb{R}^+}$.

What this simple argument shows is that the same procedure used in
\cite{R} for constructing wave functions in the energy
representation can be used to construct a rigged Hilbert space of
Hardy functions.  If, as claimed in \cite{RI}, $U_\pm$ are unitary
and the Hamiltonian is mapped by $U_\pm$ to the multiplication
operator in $L^2({\mathbb{R}}^+,dE)$, then all of the conditions
enumerated in Section \ref{sec2}, in particular
\eqref{2.7}--\eqref{2.12}, are fulfilled for the spherical shell
potential used in \cite{R} and thus a
 rigged
Hilbert space of Hardy functions can be built for this potential. In
this sense, albeit inadvertently, \cite{R} has given us an explicit
 method of constructing the
spaces $\Phi_\pm$ in the position representation that leads to the
rigged Hilbert spaces \eqref{2.1} (or, equivalently, \eqref{2.7})
for the spherical shell potential \eqref{3.2}: choose
$\Phi_\pm:=U_\pm^{-1}\left[\left.\S\cap\H_\pm^2\right|_{\mathbb{R}^+}\right]$,
where $U_\pm$ are the transformations defined by \eqref{3.16} and
\eqref{3.17}. The method runs closely parallel to the general
procedure outlined in \eqref{2.8} for constructing rigged Hilbert
spaces of Hardy functions. The only remaining problem here is to
identify the form and properties of those functions in
$L^2({\mathbb{R}}^+,dr)$ that are in one-to-one correspondence with
the subspaces $\left.\S\cap\H_\pm^2\right|_{\mathbb{R}^+}$ by way of
the unitary transformations $U_\pm$. While this requires further
mathematical investigations, the main features of the rigged Hilbert
spaces of Hardy class functions and the properties of the resulting
theory of resonance scattering and decay for the spherical shell
potential have been exemplified. In other words, the example studied
in \cite{R} answers  its own complaint ``nobody has found a
potential to which such a [time asymmetric quantum] theory applies''
in the affirmative. The article \cite{R} misses this conclusion as a
result of mathematical mistakes.

\section{On other aspects of the critique}\label{sec4}

The main weakness of \cite{R} is the rather rudimentary mathematical
mistakes that we have outlined in Section \ref{sec3}. In view of
these mathematical errors,  the crux of the case of \cite{R} against
rigged Hilbert spaces of Hardy functions falls apart. However, the
article is also riddled with erroneous remarks as well as citations
attributed to the proponents of TAQT. In the following subsections,
we will point out some of the more glaring errors of diffused focus
of \cite{R} with respect to citing from the literature and errors
 of
reasoning with respect to drawing strong physical conclusions from
mathematical fallacies.
\begin{itemize}
\item[{4.1.}]\label{item4.1}
After an inaccurate, extended and partially misleading paraphrasing
of the section 3 of \cite{BAK}, the author of \cite{R} writes on
page 9262:
\begin{enumerate}
\item[{}] ``Then, as a mathematical statement for `no preparations
for $t>0$', one takes
 $$
    |\langle E|\varphi^{\rm in}(t)\rangle|=|\langle
    ^+E|\varphi^+(t)\rangle|=0\,,\qquad t>0\eqno{(4.3)\ \text{of}\
\cite{R}}
$$
for all energies, which implies
$$
    0=\int_{-\infty}^\infty dE\, \langle E|\varphi^{\rm in}(t)\rangle
    = \int_{-\infty}^\infty dE\,\langle
    ^+E|\varphi^+(t)\rangle= \int_{-\infty}^\infty dE\, \langle
    ^+E|e^{-itH}|\varphi^+\rangle\eqno{(4.4)\ \text{of}\ \cite{R}}
$$
{\it or}
$$
0= \int_{-\infty}^\infty dE\,
e^{-itE}\,\varphi^+(E)\equiv\widetilde\varphi^+(t)\,, \qquad {\rm
for}\quad t>0 \eqno{(4.5)\ \text{of}\ \cite{R}{\rm "}}
$$
\end{enumerate}
\noindent Then, the author of \cite{R} goes on to conclude:
\begin{enumerate}
\item[{}] ``The first flaw lies in assumption (4.3). From such an
assumption, it follows that
$$
0= \langle E^+|\varphi^+(t)\rangle
=e^{-itE}\,\varphi^+(E)\eqno{(4.9)\ \text{of}\ \cite{R}}
$$
for all energies. Hence,
$$
0=\varphi^+(E)\eqno{(4.10)\ \text{of}\ \cite{R}}
$$
for all energies, which can happen only when $\varphi^+$ is
identically $0$. Thus, the preparation-registration arrow of time
holds only in the meaningless case of the zero wavefunction.''
\end{enumerate}

In contrast, in the original article \cite{BAK}, almost the same
words are used to justify another mathematical statement:
\begin{enumerate}
\item[{}] ``As the mathematical statement for ``no preparations
for $t>0$'' we therefore write (the slightly weaker condition)
\begin{equation}
    0=\int dE\,\langle E|\phi^{\rm in}(t)\rangle =\int
    dE\langle^+E|\phi^+(t)\rangle =\int
    dE\,\langle^+E|e^{-itH}|\phi^+\rangle\nonumber
    \end{equation}
{\it or}
$$
0=\int_{-\infty}^\infty dE\,
\langle^+E|\phi^+\rangle\,e^{-itE}\equiv {\cal F}(t)\qquad {\rm for}
\quad t>0\eqno{(3.4)\ \text{of}\ \cite{BAK}{\rm "}}
$$
\end{enumerate}

That is, the authors of \cite{BAK} never use the `assumption' (4.3)
of \cite{R} anywhere in their derivation. In particular, they do not
use the vanishing of the integrand to conclude that integral (3.4)
of \cite{BAK} vanishes, as the author of \cite{R} (mis)leads his
readers into inferring. It is (3.4) of \cite{BAK} (or (4.4) of
\cite{R}) that the authors take as the mathematical statement for
``no preparations for $t\geq0$'' and not (4.3) of \cite{R}.

While it is certain that the assumption $\langle E, \eta|\phi^{\rm
in}(t)\rangle=0$ for $t>0$ (or, $\langle E^+,
\eta|\phi^+(t)\rangle=0$ for $t>0$) implies $\phi^{\rm in}=0$ (or
$\phi^+=0$), it is no less certain that (3.4) of \cite{BAK} does not
imply (4.10) of \cite{R}. According to the Paley-Wiener theorems
\cite{K,KI}, the mathematical statement (3.4) of \cite{BAK} for ``no
preparations for $t\geq0$'' is equivalent to the hypothesis
 that
$\langle E|\phi^{\rm in}\rangle=\langle^+E|\phi^+\rangle$ is a Hardy
function in the open lower half plane $\{z:\ z\in{\mathbb C};{\rm
Im\,}z<0\}$. This equivalence  has been extensively discussed in
many other references, including \cite{BMLG,BI,BGMA,BGM,REL,BLV}.

It appears at a first glance that an objection can be made, as the
author of \cite{R} does,  to the fact that integration over energy
in (3.4) of \cite{BAK} extends from $-\infty$ to $+\infty$ rather
than from $0$ to $\infty$, the physical spectrum of energy. However,
we point out here that (3.4) of \cite{BAK} applies to the space
$\H_-^2$ (more precisely, the subspace $\S\cap\H_-^2$) and not
 the
physical space $\left.\S\cap\H_-^2\right|_{\mathbb{R}^+}$.
  In particular, the vanishing of the integral
(3.4) of \cite{BAK} establishes, by way of Paley-Wiener theorems,
that the operator $-i\E$ generates a one parameter semigroup
$e^{-i\E t}$, $t\geq0$, in $\H_-^2$. From this and the topological
properties of the space $\S\cap\H_-^2$ it follows that $e^{-i\E t}$
is a differentiable semigroup in $\S\cap\H_-^2$. Now, by van
Winter's theorem \cite{VW} it finally follows that $e^{-i\E t}$ is a
differentiable semigroup in the space
$\left.\S\cap\H_-^2\right|_{{\mathbb{R}}^+}$. It is this space
$\left.\S\cap\H_-^2\right|_{{\mathbb{R}}^+}$ consisting of functions
defined over the physical energy spectrum $[0,\infty)$ that we take
as the space of `in' wavefunctions and the time evolution of these
wavefunctions is given by a Hamiltonian generated differentiable
semigroup. A similar computation shows that $e^{i\E t}$, $t\geq0$,
is a differentiable semigroup defined in
$\left.\S\cap\H_+^2\right|_{\mathbb{R}^+}$, the space of functions
that represent the measured `out'  states (more properly called
out-observables).

Thus, the author of \cite{R} is mistaken about both the physics and
mathematics of \cite{BAK}. In our view, the only objection that can
be made against \cite{BAK} is semantic in nature; namely, the
condition (3.4) of \cite{BAK} should not be called ``slightly
weaker'' than the condition $\<E,\eta^+|\phi^+(t)\>=0$ for it is a
mathematically different condition, equivalent to the Hardy
hypothesis.

\item[{4.2.}]\label{item4.2}
Section 5 of \cite{R} is simply a reiteration of its Section 3.
Here, the author of \cite{R} claims again that the wavefunctions
for the spherical shell potential cannot be of Hardy class because
they are not square integrable when extended to negative energy
values, in contrast to Hardy class functions which are square
integrable over the whole real line. To ``prove'' the non-square
integrability of the energy wavefunctions over the negative
semi-axis, the author uses the inequality ((5.11), \cite{R})
\begin{equation}
\left|\phi^+(E)\right|\leq
C\frac{\left|E\right|^{1/4}}{1+\left|E\right|^{1/2}A}e^{\left|E\right|^{1/2}A}\label{4.10}
\end{equation}
and makes the now familiar conclusion that $\left|\phi^+(E)\right|$
blows up for $E\rightarrow-\infty$ simply because it is bounded from
above by a divergent function.

The entire Section 5 of \cite{R} is written in a rather contrived
 manner.
The notation $(\varphi^+,\varphi^+)_{\rm Hardy}$ and equations
(5.15) and (5.16) suggest that Hardy class functions are not square
integrable and that TAQT suffers from these divergences. It is
standard  knowledge that Hardy class functions $\H_\pm^2$ are
subspaces of the Hilbert space $L^2({\mathbb{R}})$ \cite{K}
 and thus equations (5.15) and (5.16) of \cite{R} do not hold for Hardy
 functions,
despite the author's notation that suggests so.

The last paragraph of Section 5 of \cite{R} is laden with mangled
logic and mathematical fallacies. For instance, with the assertion
``{The problem is that, although there is nothing wrong with
analytically continuing from the physical spectrum into the negative
real line of the second sheet, the resulting integrals are not
convergent, as shown above}",  the author of \cite{R} lays the
charge against the Hardy space theory with the `problem' of
divergence of his own wavefunctions, itself a non sequitur in view
of \eqref{4.10}. To reiterate, Hardy functions are perfectly square
integrable over the whole real line. Therefore, the integral (5.13)
of \cite{R} converges if Hardy functions are used in the integrand.
Likewise, if $\varphi^+$ belongs to the space
$\left.\S\cap\H_-^2\right|_{\mathbb{R}^+}$,  as required in TAQT,
then the integral (5.14) is also convergent. Therefore, the
divergences that the author refers to, in particular (5.13)-(5.16),
first of all, are wrong in view of \eqref{4.10} and our explanation
of the error just below it in the text, and second of all have
nothing to do with Hardy functions.
 In other words,
even if one accepts (5.13) and (5.14) of \cite{R}, these would apply
to the wavefunctions of \cite{R} and not to Hardy functions.

The author of \cite{R} has certainly proven no mathematical
statement about the semiboundedness of the Hamiltonian, other than
simply stating his opinion that Hardy functions are not consistent
with this requirement. Once again, let us restate what we stated in
Section \ref{sec2}: The Hilbert space $L^2({\mathbb{R}}^+, dE)$
consists of square integrable functions defined over the {\em
positive real line} $\mathbb{R}^+$. The spaces
$\left.\S\cap\H_\pm^2\right|_{\mathbb{R}^+}$ are subspaces of
$L^2({\mathbb{R}}^+, dE)$, and thus the functions of
$\left.\S\cap\H_\pm^2\right|_{\mathbb{R}^+}$ have domains
$[0,\infty)$. The Hamiltonian, which has the realization as the
multiplication operator in $L^2({\mathbb{R}}^+, dE)$ with the domain
$\left.\S\cap\H_+^2\right|_{\mathbb{R}^+}$ or
$\left.\S\cap\H_-^2\right|_{\mathbb{R}^+}$, is essentially
self-adjoint and has a non-degenerate, absolutely continuous
spectrum that coincides with the positive real line $\mathbb{R}^+$.
These facts have been well established not only for the
non-relativistic case considered in \cite{R} but also for the
relativistic case \cite{REL}. The author of \cite{R} does nothing by
way of showing where the proponents of TAQT have made mathematical
errors. In fact, to quote the author of \cite{R}, ``{\em it is not
that the math of Bohm-Gadella theory is wrong, it is rather that the
math of the Bohm-Gadella theory is inconsistent with quantum
mechanics''. } However, the assertion that Hardy hypothesis is
inconsistent with the semiboundedness of the Hamiltonian (\cite{R},
page 9265, last paragraph) is inexorably a mathematical statement
and it is not in the spirit of mathematical physics to make such a
definitive pronouncement without even an attempt at a proof.

As a final remark on Section 5 of \cite{R}, let us make believe that
the author of \cite{R} has not made mathematical mistakes and has
found a set of wavefunctions that have the divergence properties
described in \cite{R}. Then the only conclusion that one can draw
from the study of \cite{R} is that there are solutions to the
Schr\"odinger equation
 that are not of Hardy class. This does not provide a basis for
determining if TAQT is mathematically or physically tenable. Nobody
has ever asserted that {\em all} solutions of the
 the Schr\"odinger equation are of Hardy class,
in much the same way that nobody has ever asserted, for instance,
that all solutions of the Schr\"odinger equation are square
integrable. It is common knowledge that the Schr\"odinger equation
must be solved subject to some physically meaningful boundary
conditions, such as square integrability. In TAQT, the Hardy
hypothesis constitutes such a boundary condition subject to which
the quantum dynamical equations must be solved. Finding solutions
which do not obey these boundary conditions means just that: there
exist solutions that do not obey the boundary conditions of TAQT.
The existence of other solutions is not tantamount to a
counter-example that invalidates TAQT. In fact, the author of
\cite{R} does not even succeed
 in
showing that there are non-Hardy solutions in his example, much less
prove that there are no Hardy solutions.

\noindent{In connection with Section 6 of \cite{R}, we wish to make
the following remarks:}

\item[{4.3.}]
Hardy spaces  do not appear to be essential for the construction of
vectors with complex energy eigenvalues that represent resonances.
In fact, what really matters for the construction of these vectors
is the use of wavefunctions that, in the energy representation,
admit suitable analytic extensions. The subspace of these
wavefunctions should be dense in the whole Hilbert space of states.

That we can use different kinds of spaces of analytic functions can
be shown as follows. Let us consider infinitely differentiable
functions with compact support. By Paley-Wiener theorems, their
Fourier transforms are entire analytic functions \cite{RU}. The
values these functions on the positive semiaxis determine their
values on the whole complex plane by the Principle of Analytic
Continuation \cite{MA}. Moreover, the restrictions of these entire
functions to $\mathbb{R}^+$ form a dense subspace of $L^2({\mathbb
R}^+, dE)$. This dense subspace can be used to obtain a class of
resonance state vectors.

Hardy functions are used in TAQT for the following reasons:
\begin{enumerate}
\item[{(a)}]
The intersection of Hardy functions with Schwartz functions fulfills
an adequate set of mathematical conditions for a consistent theory
of scattering and decay. In particular, as outlined in Section
 \ref{sec2}, the
spaces $\left.\S\cap\H_\pm^2\right|_{{\mathbb{R}}^+}$ lead to the
rigged Hilbert spaces \eqref{2.1}. The decaying Gamow vectors can be
properly defined as elements in the dual space
$\left(\left.\S\cap\H_\pm^2\right|_{{\mathbb{R}}^+}\right)^\times$.

\item[{(b)}] The rigged Hilbert spaces of Hardy functions \eqref{2.1}
permit a time asymmetric formulation of resonance phenomena
\cite{AP,BMLG,BI,BGM,BGMA,BLV, AGPP, AGMP,REL,CGL}

 \item[{(c)}] Functions that are both Hardy and Schwartz have
very good behavior at infinity, both on the real line and on the
appropriate complex semi-plane. This regularity property can be
used, for instance, to give a workable formula for the background
that is always present in all resonance processes as an integral
over the negative semiaxis. A further advantage of Hardy functions
is that by means of the Mellin transform the background integral can
be written as an integral over the positive semiaxis \cite{GII}.
\end{enumerate}

\item[{4.4.}] It is inaccurate to state that the proponents of TAQT
dispense with the principle of asymptotic completeness. Asymptotic
completeness means the  `in' scattering states and
 `out' scattering states inhabit the same Hilbert space, i.e., the
 ranges of
the M{\o}ller operators are the same, $\H_{\rm in}=\H_{\rm out}$.
What TAQT establishes is that the `in' and `out' scattering vectors
inhabit two different vector spaces $\Phi_-$ and $\Phi_+$, each of
which is dense in the same Hilbert space. Therefore, the Hilbert
space completion of $\Phi_+$ and $\Phi_-$ gives the conventional
principle of asymptotic completeness and thus TAQT is not in
contradiction with conditions that define the M{\o}ller operators or
the $S$-operator. In fact, as we have seen in Section 2, M{\o}ller
operators play a role in the construction of the rigged Hilbert
spaces of Hardy functions \eqref{2.1}, and asymptotic completeness
has been used in the early publications on TAQT \cite{GI} as well as
later review articles \cite{BG,CG}. TAQT does introduce a
topological refinement to the principle of asymptotic completeness
in that $\Phi_+\not=\Phi_-$ (both as dense subspaces of $\H$ and as
complete topological vector spaces), and this is the content of the
statements that the author attributes to the proponents of TAQT.

 \item[{4.5.}]
No serious questions exist about the construction of Gamow vectors
by solving the Schr\"odinger equation subject to  purely outgoing
boundary conditions. In fact, there is a well-known procedure to
obtain Gamow vectors this way \cite{N,CG,BAUM}.

\item[{4.6.}]
To conclude this section, we want to comment on the last assertion
of \cite{R}: {\it The major achievement of the Bohm-Gadella theory,
namely time asymmetry and the rigorous construction of resonance
states, can be achieved rigorously within standard quantum
mechanics}. The author does not give any references to support his
assertion and it is not clear what the author means by `standard
quantum mechanics.'  If by `standard quantum mechanics' one means
the use of a  Hilbert space to represent the space of states, we are
aware of two mathematically
 rigorous
approaches to resonances. One is given by the Lax-Phillips theory
\cite{SHE,S,P}. The other is the dilation of analytic potentials or
complex scaling \cite{RSIV}. The TAQT has some features that are
found also in these formalisms but it has other strengths that make
it rather unique.

The Lax-Phillips formalism is attractive and, in principle,
 it accommodates time asymmetry. It works in the
context of the Hilbert space but the price to pay for retaining the
Hilbert space is that Gamow vectors  are eigenvectors of a non
self-adjoint, dissipative operator with complex eigenvalues. In
contrast,  in TAQT (based on rigged Hilbert spaces of Hardy
functions)  Gamow vectors are eigenvectors of an extension (into the
dual space $\Phi_+^\times$) of a self-adjoint Hamiltonian. In both
formalisms the eigenvalues are the same and coincide with the
resonance poles of the analytic $S$-matrix. Furthermore, the two
formalisms are closely connected as shown by Y.~Strauss \cite{S}.

In the complex scaling formalism, the Hamiltonian is ``dilated'' by
a complex parameter and the dilated Hamiltonian is no longer
self-adjoint. The relevant part of this complex parameter is its
imaginary part that behaves like an angle. The number of resonances
defined as complex eigenvalues of the dilated Hamiltonian is
directly related to the size of the imaginary part of the dilation
parameter but the resonance parameters themselves do not depend on
the value of the dilation parameter. Moreover, the resonance
eigenvalues of the dilated Hamiltonian coincide with the poles of
the extended resolvent.
\end{itemize}

\section{Concluding remarks.}

We agree with the sentiment expressed by the title of \cite{R}: if
by `standard quantum mechanics' the author means unitary evolutions
and Hilbert spaces, indeed TAQT is not consistent with `standard
quantum mechanics'. Aside from this, as we have shown in the
foregoing response,  \cite{R} does not bear scrutiny as a critique
on  Time Asymmetric Quantum Theory
 for two
reasons: first, the author makes  mistakes that the main
mathematical conclusions of \cite{R} are false; second, even if
there exist proper mathematical methods by which one can arrive
 at solutions
to the Schr\"odinger equation for the spherical shell potential that
are not of Hardy class, this is not equivalent to saying that there
are no solutions to the Schr\"odinger  equation that are of Hardy
class for this potential or, much less, all other. Therefore, even
if it had been conducted with proper mathematical care, the study of
\cite{R} could not have led to a conclusion about the veracity TAQT.
While critical inquiry is crucial to the practice and philosophy of
science, the one carried out in \cite{R} falls short of the
normative technical rigor and reasoning of theoretical physics that
it does not inform.

\section*{Acknowledgements.} M.G.~acknowledges the financial support
from the Junta de Castilla y Le\'on Project VA013C05 and the
Ministry of Education and Science of Spain, projects MTM2005-09183
and FIS2005-03988. S.W.~acknowledges the financial support from the
University of Valladolid where he was a visitor while this work was
done and additional financial support from Grinnell College.

\end{document}